\documentclass[twocolumn]{aastex63}
\usepackage{graphicx}
\usepackage{amsmath}
\usepackage{multirow}
\usepackage{lineno}
\usepackage[figuresleft]{rotating}

\newcommand{\as}{^{\prime \prime}}
\newcommand{\kms}{\rm km~s^{-1}}
\newcommand{\nthp}{\rm N_{\rm 2}H^{\rm +}}
\newcommand{\hcop}{\rm HCO^{\rm +}}
\newcommand{\ceo}{\rm C^{\rm 18}O}

\received{May 2, 2022}
\revised{July 27, 2022}
\accepted{August 15, 2022}

\shorttitle{Evolution and Energy Balance of HFSs in IC~5146}
\shortauthors{Chung et al.}

\begin{document}

\title{Evolution of the Hub-filament Structures in IC~5146 in the Context of the Energy Balance of Gravity, Turbulence, and Magnetic Field}

\author{Eun Jung Chung}
\affiliation{Department of Astronomy and Space Science, Chungnam National University, Daejeon, Republic of Korea}

\author{Chang Won Lee} \affiliation{Korea Astronomy and Space Science Institute, 776 Daedeokdae-ro, Yuseong-gu, Daejeon 34055, Republic of Korea} \affiliation{University of Science and Technology, Korea (UST), 217 Gajeong-ro, Yuseong-gu, Daejeon 34113, Republic of Korea}

\author{Woojin Kwon} \affiliation{Department of Earth Science Education, Seoul National University, 1 Gwanak-ro, Gwanak-gu, Seoul 08826, Republic of Korea} \affiliation{SNU Astronomy Research Center, Seoul National University, 1 Gwanak-ro, Gwanak-gu, Seoul 08826, Republic of Korea}

\author{Hyunju Yoo} \affiliation{Department of Astronomy and Space Science, Chungnam National University, Daejeon, Republic of Korea}

\author{Archana Soam} \affiliation{Indian Institute of Astrophysics, Kormangala (IIA), Bangalore 560034, India}

\author{Jungyeon Cho} \affiliation{Department of Astronomy and Space Science, Chungnam National University, Daejeon, Republic of Korea}

\begin{abstract}
We present the results of 850~$\mu$m polarization and $\ceo~(3-2)$ line observations toward the western hub-filament structure (W-HFS) of the dark Streamer in IC~5146 using the James Clerk Maxwell Telescope (JCMT) SCUBA-2/POL-2 and HARP instruments. We aim to investigate how the relative importance of the magnetic field, gravity, and turbulence affects core formation in HFS by comparing the energy budget of this region. We identified four 850~$\mu$m cores and estimated the magnetic field strengths ($B_{\rm pos}$) of the cores and the hub and filament using the Davis-Chandrasekhar-Fermi method. The estimated $B_{\rm pos}$ is $\sim$80 to 1200~$\mu$G. From Wang et al., $B_{\rm pos}$ of E-47, a core in the eastern hub (E-hub), and E-hub were re-estimated to be 500 and 320~$\mu$G, respectively, with the same method. We measured the gravitational ($E_{\rm G}$), kinematic ($E_{\rm K}$), and magnetic energies ($E_{\rm B}$) in the filament and hubs and compared the relative importance among them. We found that an $E_{\rm B}$-dominant filament has $aligned$ fragmentation type, while $E_{\rm G}$-dominant hubs show $no$ and $clustered$ fragmentation types. In the $E_{\rm G}$ dominant hubs, it seems that the portion of $E_{\rm K}$ determines whether the hub becomes to have $clustered$ (the portion of $E_{\rm K}\sim20\%$) or $no$ fragmentation type ($\sim10\%$). We propose an evolutionary scenario for the E- and W-HFSs, where the HFS forms first by the collision of turbulent flows, and then the hubs and filaments can go into various types of fragmentation depending on their energy balance of gravity, turbulence, and magnetic field. 
\end{abstract}


\section{Introduction}

Stars are known to mainly form in the dense clumps/cores which have been developed in filamentary molecular clouds \citep[e.g.,][]{andre2010}. Hence, painstaking efforts have been made to find how filaments and dense cores form and evolve \citep[e.g.,][]{arzoumanian2019,chung2019}. Though being still under debate in details, it is suggested that the molecular filaments first form by the dissipation of large-scale turbulence and then the dense clumps/cores are generated in the gravitationally supercritical filaments via fragmentation \citep{andre2014}. 

\begin{figure*} \epsscale{1.17}
\plotone{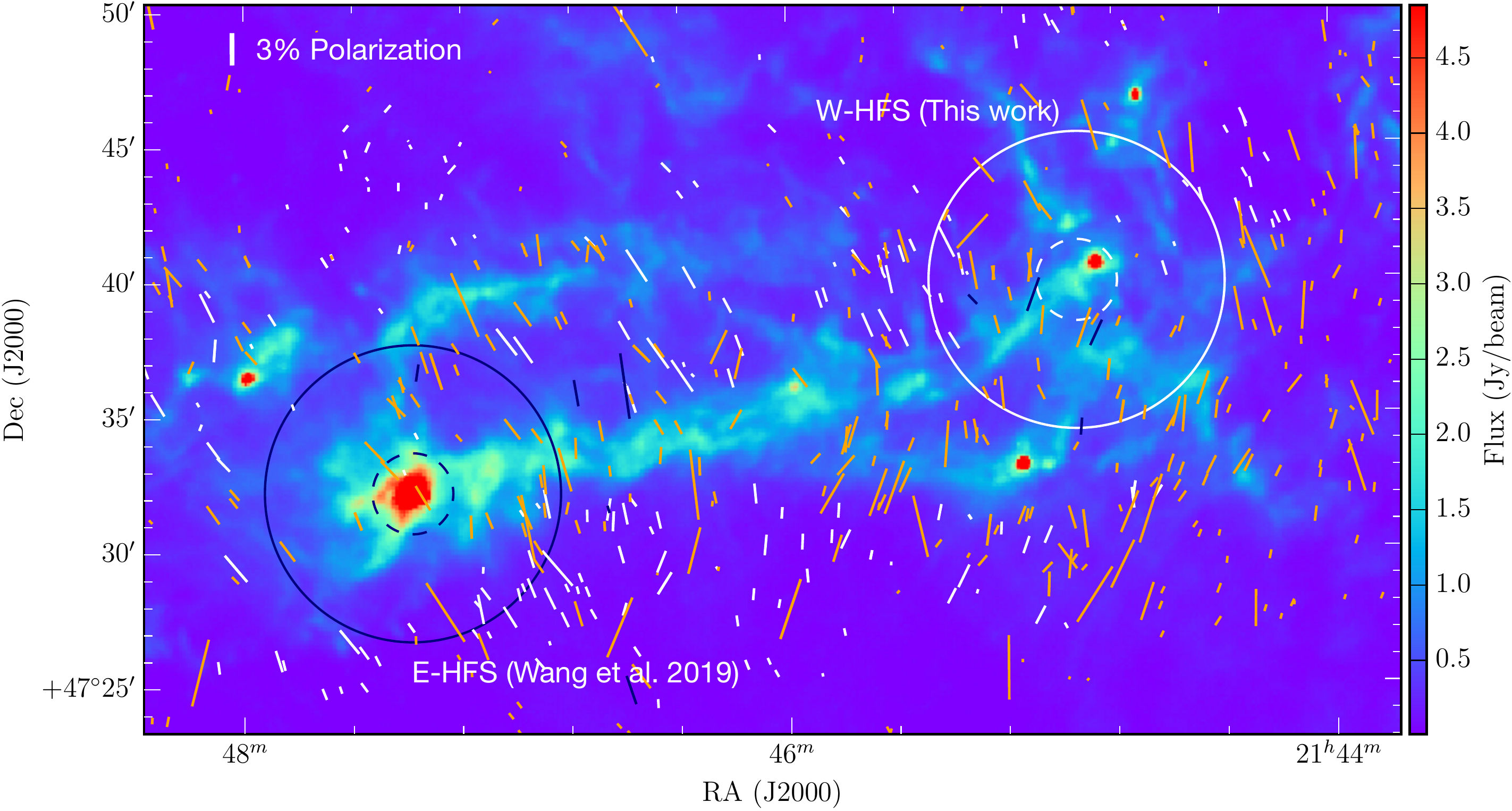}
\caption{The dark Streamer of IC~5146 with previous optical and infrared polarizations. The JCMT/POL-2 observing areas for the eastern HFS \citep[E-HFS;][]{wang2019} and the western HFS (W-HFS; this study) are indicated with navy and white circles, respectively, on the Herschel 250~$\mu$m image. The dashed circles indicate the central 3$^\prime$ region with a high sensitivity. The white, orange, and navy lines show the polarization vectors detected in AIMPOL $R_{c}$ band (0.67~$\mu$m), and Mimir $H$- (1.6~$\mu$m) and $K$-band (2.2~$\mu$m) \citep{wang2017}. \label{fig:obsregion}}
\end{figure*}

The magnetic field is generally considered to play a key role in star formation. In the molecular filament on pc scales, the striations of filaments are observed to be parallel to the magnetic field, and it is suggested that the magnetic field plays a role as an aisle for material to flow along them onto main filaments till the filaments accrete sufficient mass to collapse by gravity and form cores \citep[e.g.,][]{palmeirim2013}. In the stage of gravitational collapse, theoretically, the magnetic field is expected to provide significant support against the collapse under the self-gravity to explain the longer lifetime of molecular cloud than their free-fall collapse time \citep[e.g.,][and references therein]{mckee2007}. However, the hourglass morphology of the magnetic fields is frequently found in massive star-forming clouds, indicating that the magnetic field can be modified by gravity or outflows in sub-parsec scales \citep[e.g.,][]{wang2019,lyo2021}. Besides, the shocks from outflows, stellar feedback of expanding ionization fronts of {\sc H~ii} region, and gas flow driven by gravity are considered to cause the magnetic field distortions \citep[e.g.,][]{hull2017b,pillai2020,arzoumanian2021,eswar2021}. Hence, the significance of magnetic field may change from time to time as well as from cloud to cloud. In this paper, we replace the question, how stars form, into that how gravity, turbulence, and magnetic field play roles in forming stars, especially in the process of the fragmentation from cloud to clumps/cores.

The precise roles of the gravity, turbulence, and magnetic field are still unclear, especially at the different evolutionary stages of filaments and dense cores. The clouds and star formation models suggest that the magnetic field and turbulence may have different importance in the evolution stages \citep[e.g.,][]{crutcher2012}. Moreover, the subtle difference in the relative significance among the gravity, turbulence, and magnetic field can make different evolution from the clump to the core scale \citep[e.g.,][]{hennebelle2011,tang2019}. 

In this paper, we investigate the roles of gravity, turbulence, and magnetic field of the western hub-filament structure (W-HFS) of the dark Streamer in IC~5146. Hub-filament structures (HFSs), consisting of a central hub with relatively higher column density ($> 10^{22} ~\rm cm^{-2}$) and several filaments extended from the hub with relatively low column density and high aspect ratio, are easily found in nearby star-forming molecular clouds and more distant infrared dark clouds \citep{myers2009}. The central hub of HFS is frequently observed to be associated with stars and stellar clusters and thus likely a birth place of stellar clusters \citep[e.g.,][]{gutermuth2009,myers2009,kumar2020}. Therefore, HFSs of nearby star-forming molecular clouds may be one of the best laboratories to test the initial conditions in the formation of stars and stellar clusters.

The dark Streamer of IC~5146 locates on the northwest of the Cocoon Nebula, in the constellation Cygnus. It has a long filamentary shape, and two prominent HFSs locate in the eastern and the western parts as shown in Figure~\ref{fig:obsregion}. The properties of filaments and dense cores in IC~5146 are studied with various molecular lines as a part of the ``TRAO survey of the nearby Filamentary molecular clouds, the Universal Nursery of Stars" \citep[TRAO FUNS;][]{chung2021}. Velocity coherent filaments in the IC~5146 region were identified using the 3-dimensional information of $\ceo~(1-0)$ line data which have 49$^{\as}$ spatial resolution (corresponding to 0.14~pc at the distance of 600~pc) and 0.1~$\kms$ channel width. It was found that there is a velocity coherent filament, referred to as F4 hereafter, over the dark Streamer of IC~5146, and smaller filaments and clumps with different velocities from that of F4 are overlapped in the line-of-sight direction. F4 is gravitationally supercritical, and interestingly there are two hubs in the eastern- and western-end, which are named as E-hub and W-hub, respectively. Dense cores were identified with the $\nthp~(1-0)$ data for which spatial resolution and velocity channel width are 52$^{\as}$ and 0.06~$\kms$, respectively. E- and W-hub regions are found to have one dense core each. The $\ceo~(1-0)$ line which traces the filament gas material reveals that the two hubs are supersonic ($\sigma_{\rm NT} / c_{\rm s} \sim 3$), but the $\nthp~(1-0)$ line which traces the dense cores shows that the core material is less turbulent than the filament gas ($\sigma_{\rm NT} / c_{\rm s} \sim 2$) \citep{chung2021}. 

Polarization observations of IC~5146 made with Planck show that the magnetic field is nearly uniform and prefers a perpendicular orientation to the gas column density contours \citep{planck35}. The optical and near-infrared (NIR) polarization observations toward the whole IC~5146 region present uniform magnetic field vectors perpendicular to the dark Streamer \citep{wang2017}. The submillimeter polarization observations are made toward the E-hub as a part of the BISTRO survey \citep[][W2019 hereafter]{wang2019}, showing a curved magnetic field morphology which implies the possible modification of magnetic field by gravitational contraction in the hub. We adopt the distance of IC~5146 in this study as 600$\pm$100~pc measured by \citet{wang2020a} using GAIA DR2. 

The paper is organized as follows. In Section~\ref{sec:obsdr}, we describe the observation and data reduction. In Section~\ref{sec:results}, we present the results of the observations and the measured magnetic field strength. We analyze and discuss implications on our results in Section~\ref{sec:anal} and \ref{sec:disc}, respectively, and summarize all the main results in our study in Section~\ref{sec:summ}. \\

\section{Observations} \label{sec:obsdr}

\subsection{Polarization Observations}

The western HFS of IC~5146 was observed with the SCUBA-2/POL-2 instrument on the James Clerk Maxwell Telescope (JCMT) between 2020 June 17 and July 8. The region was observed 21 times, and each data set has an average integration time of 41 minutes at the weather band~2 ($0.05 < \tau_{\rm 225~GHz} \leq 0.08$). The observations are made by POL-2 daisy map mode which covers a circular region of 11$^\prime$ diameter with the best sensitivity coverage of central 3$^\prime$ of the map. SCUBA-2/POL-2 simultaneously obtains the data at 450~$\mu$m and 850~$\mu$m wavelengths with the effective beam sizes of 9.6$^{\as}$ and 14.1$^{\as}$ (0.028 and 0.041~pc at a distance of 600~pc), respectively. We present the results from the 850~$\mu$m data only in this paper.

The 850~$\mu$m data were reduced using the STARLINK/SMURF package {\tt pol2map}. The data reduction process follows three main steps. In the first step, the raw bolometer time streams for each observation are converted into separate Stokes {\em I}, {\em Q}, and {\em U} time streams using the process {\it calcqu}. Then, an initial Stokes {\em I} map is created for all observations via the iterative map-making process {\it makemap}. In the second step, with the initial {\em I} map, a mask is iteratively determined based on the signal-to-noise ratio, and the background pixels defined by the mask are set to zero at the end of each iteration within {\it makemap}. The use of a mask produces an improved individual {\em I} map by preventing the growth of gradients and any artificial large scale structure, and protecting the various noise models' evaluation from bright sources. The final {\em I} map is produced by co-adding the improved individual {\em I} maps. In the final step, {\em Q} and {\em U} maps are created from the {\em Q} and {\em U} time streams with the same masks used in the previous step. The instrumental polarization is corrected with the final improved {\em I} map using the `August 2019' IP model \citep{friberg2018}. Then final {\em I}, {\em Q}, and {\em U} maps are produced with a pixel size of 4$^{\as}$, and the final debiased polarization vector catalog is provided with a bin-size of 12$^{\as}$ which is close to the beam size of the JCMT/POL-2 at 850~$\mu$m to increase the signal-to-noise ratios in the polarization data.

The Stokes {\em I} parameter is the total intensity of the incoming light, and the Stokes {\em Q} and {\em U} parameters are defined as:
\begin{equation}
	Q = I \times P \times \rm cos(2\phi) \label{eq:q}
\end{equation}
and
\begin{equation}
	U = I \times P \times \rm sin(2 \phi),  \label{eq:u}
\end{equation}
where $P$ is the polarization fraction and $\phi$ is the polarization angle. Because the polarized intensity {\em PI} is a form of quadratic sum of {\em Q} and {\em U}, $PI=\sqrt{Q^{2} + U^{2}}$, the noises of {\em Q} and {\em U} always make a positive contribution to the polarization intensity \citep[e.g.,][]{vaillancourt2006}. The debiased polarization intensity is estimated by the modified asymptotic estimator \citep{plaszczynski2014} as follows:
\begin{equation}
	PI = \sqrt{Q^{2} + U^{2}} - \sigma^{2} \frac{1 - e^{-(Q^{2} + U^{2})/\sigma^{2}}}{2\sqrt{Q^{2} + U^{2}}},
\end{equation}
\noindent where $\sigma^{2}$ is the weighted mean of the variances on {\em Q} and {\em U}:
\begin{equation}
	\sigma^{2} = \frac{Q^{2} \sigma_{Q}^{2} + U^{2} \sigma_{U}^{2}}{Q^{2} + U^{2}},
\end{equation}
and $\sigma_{Q}$ and $\sigma_{U}$ are the standard errors in {\em Q} and {\em U}, respectively.

The debiased polarization fraction $P$ and its corresponding uncertainty are calculated as
\begin{equation}
	P = \frac{PI}{I}  \label{eq:p}
\end{equation}
and 
\begin{equation}
	\sigma_{P} = \sqrt{\frac{\sigma^{2}}{I^{2}} + \frac{\sigma_{I}^{2}(Q^{2} + U^{2})}{I^{4}}} ,
\end{equation}
where $\sigma_{I}$ is the standard error in {\em I}.

We selected polarization measurements with the criteria that (1) the signal-to-noise ratio (S/N) of total intensity is larger than 10 ($I / \sigma_{I} > 10$) and (2) the polarization fraction is larger than 2 times of its uncertainty ($P / \sigma_{P} > 2$). 

\begin{figure*} \includegraphics[width=0.99\textwidth, height=.94\textwidth]{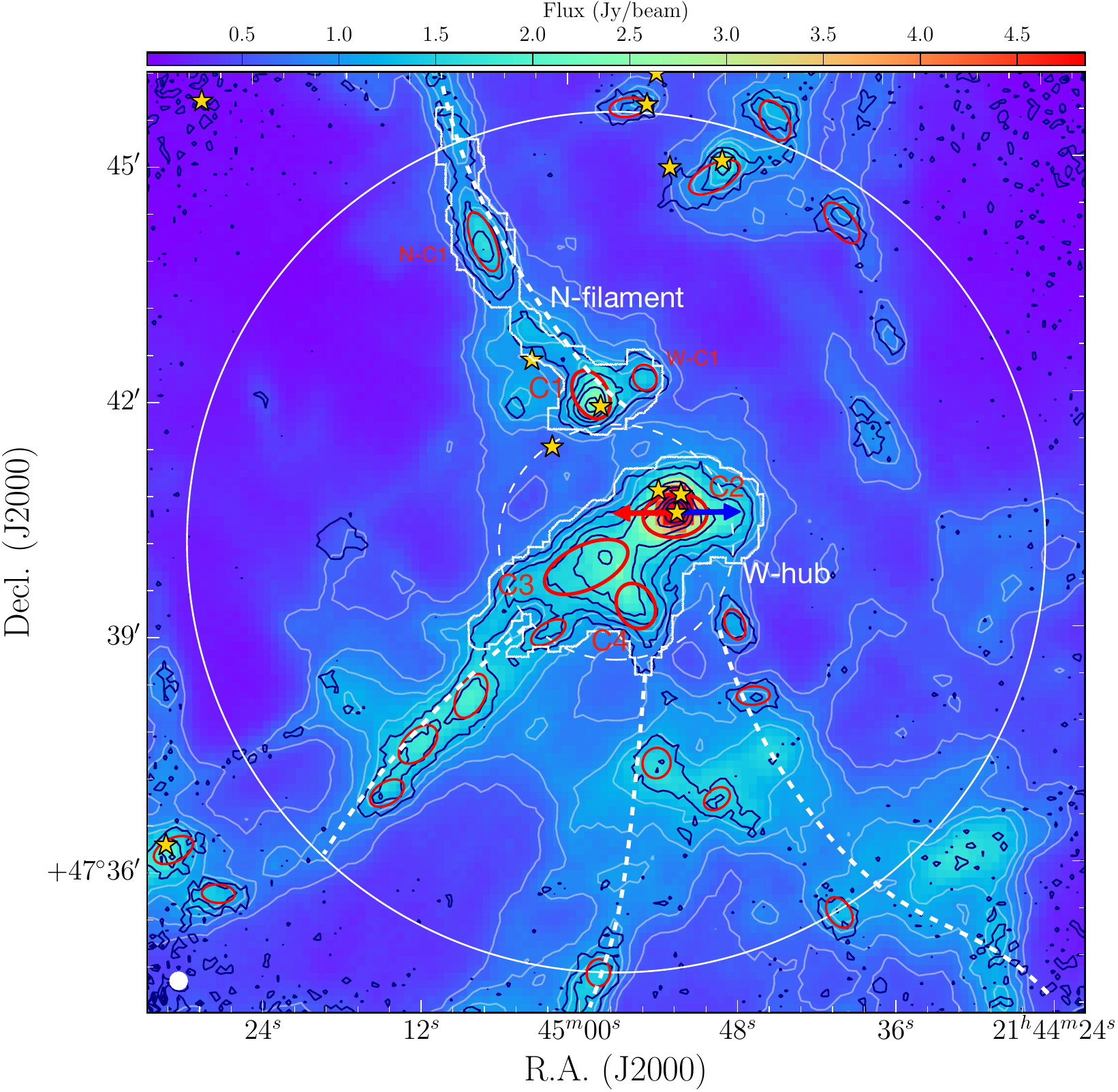}
\caption{850 $\mu$m Stokes {\em I} map (navy contours) on 250~$\mu$m Herschel map (image and white contours) with 850 $\mu$m cores (red ellipses) and W-hub and N-filament (white polygons) identified with \textsc{FellWalker} (\S~\ref{ssec:coreid}). Core number is given with red for the cores analyzed in this study. The contour levels of 850~$\mu$m emission are 3, 10, 20, 30, 50, 70, and 90$\times \sigma$ (1$\sigma =3.1$~mJy~beam$^{-1}$) and those of 250~$\mu$m emission are 6, 8, 10, and 12$\times \sigma$ (1$\sigma =0.1$~mJy~beam$^{-1}$). YSOs identified by Spitzer \citep{harvey2008} and 70~$\mu$m point sources from Herschel/PACS Point Source Catalogue \citep[HPPSC;][]{poglitsch2010} are presented with yellow stars. The red and blue arrows show the red- and blue-shifted lobes of the prominent outflow associated with the IRAS~21429+4726 \citep{dobashi2001}. The guide lines for the $\ceo$ filaments whose skeletons were extracted with the {\sc filfinder} algorithm are overlaid with the white dashed lines \citep{chung2021}. The white solid and dashed large circles indicate the full area POL-2 daisy map mode which covers a circular region of 11$^{\prime}$ diameter and smaller inner area with the best sensitivity coverage of central 3$^{\prime}$ of the map, respectively. The white circle at the bottom left corner shows the POL-2 850~$\mu$m beam size of 14.1~arcsec. \label{fig:850cores}}
\end{figure*}

The Flux-Conversion Factor (FCF) of 668~Jy~beam$^{-1}$~pW$^{-1}$ is used for the Stokes {\em I}, {\em Q}, and {\em U} data at 850~$\mu$m data in this paper. The FCF is determined by multiplying the standard 850~$\mu$m SCUBA-2 flux conversion factor 495~Jy~beam$^{-1}$~pW$^{-1}$ by 1.35 to correct the additional losses from POL-2 \citep{mairs2021}. The rms noise values in the {\em I}, {\em Q}, and {\em U} data binned to a pixel size of 12$\as$ are 3.1, 2.9, and 2.8~mJy~beam$^{-1}$, respectively.  \\

\begin{figure*} \epsscale{1.17}
\plotone{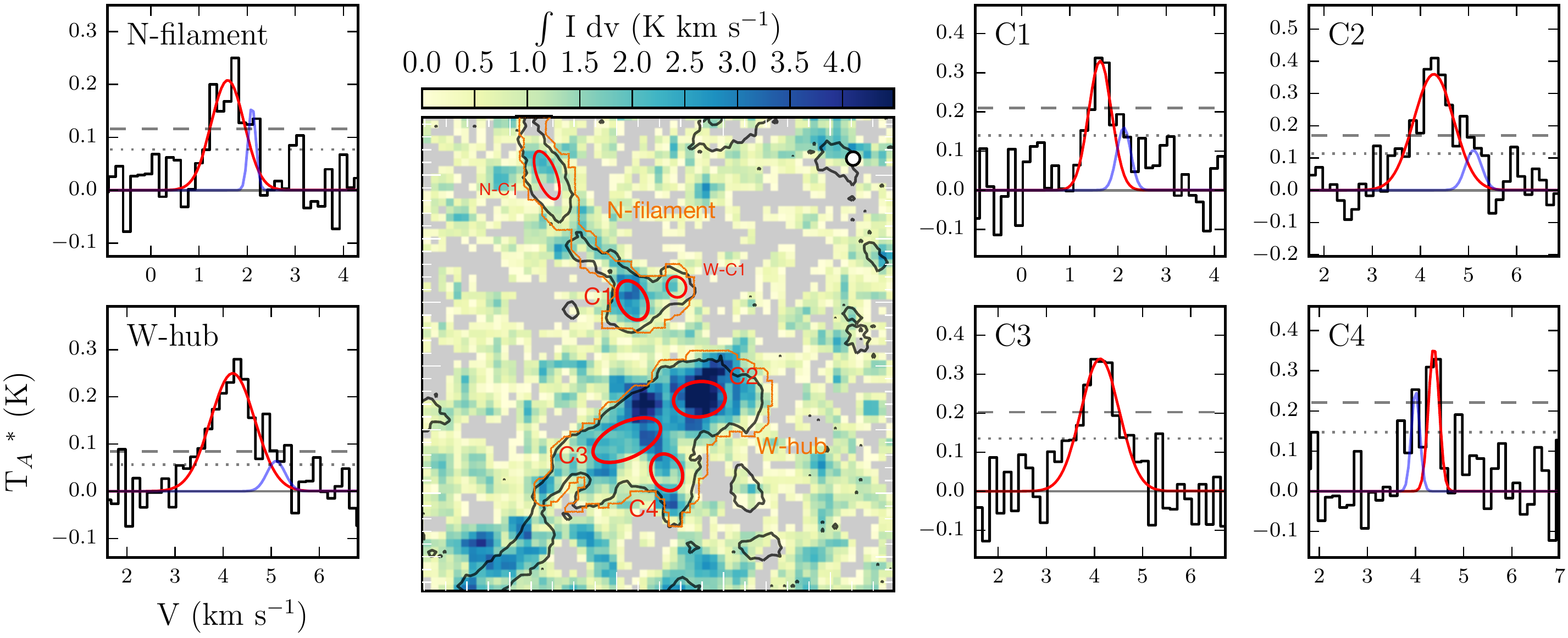}
\caption{The integrated intensity map of $\ceo~(3-2)$ (center) and the averaged $\ceo$ spectra of N-filament, W-hub, and 850~$\mu$m cores. The black contour and the red ellipses overlaid on the $\ceo$ moment 0 map indicate the 3$\sigma$ level of 850~$\mu$m emission and cores. 
Red and blue profiles overlaid on the spectra are the decomposed major and minor Gaussian components. The dashed and dotted horizontal lines indicate the 3$\sigma$ and 2$\sigma$ levels of the spectra, respectively. \label{af:c18oprof}}
\end{figure*}

\subsection{The $\ceo~(3-2)$ Line Observations} \label{sec:c18o}

\noindent To estimate the velocity dispersion of cores in the W-HFS, we carried out the $\ceo~(3-2)$ line observations toward the $9^{\prime} \times 9^{\prime}$ area of W-HFS with the Heterodyne Array Receiver Programme (HARP) on the JCMT \citep{buckle2009}. The observations have been conducted over six nights between 2020 August 28 and September 20 with the weather band~2 ($0.05 < \tau_{\rm 225~Ghz} \leq 0.08$). The data were taken in the raster mode at the default sample spacing of 7.3~arcsec. The total observing time for the $\ceo~(3-2)$ line is about 10 hours. The spatial resolution is $\sim 14^{\as}$ at 330~GHz, which is the same as that of JCMT/POL-2 850~$\mu$m data. The data are reduced using the ORAC Data Reduction (ORAC-DR) pipeline in {\sc starlink} software \citep{buckle2012}. The mean rms level is $\sim 0.06$~K in the final datacube with 7.3~arcsec pixel size and 0.15~$\kms$ channel width. \\

\section{Results} \label{sec:results}

\subsection{Identification of the 850~$\mu$m Cores} \label{ssec:coreid}

Figure~\ref{fig:850cores} shows the 850~$\mu$m Stokes {\em I} contours on Herschel 250~$\mu$m image. The 250~$\mu$m emission demonstrates well the structure of W-HFS where several elongated structures are connected with the hub. The guide lines of filaments identified with the $\ceo~(1-0)$ emission \citep{chung2021} are drawn with white dashed lines. Simply introducing the structure of W-HFS which is revealed with the 3-dimensional $\ceo~(1-0)$ datacube here, there are four filaments converging into the central hub. However, the northern filament (N-filament hereafter) is well separated from the W-hub in velocity dimension ($\Delta v \sim 2~\kms$) while the other three filaments are connected in the position-position-velocity space. Hence, the W-HFS consists with the W-hub and the three southern filaments, while the N-filament seems physically well separated from the W-hub.

\begin{deluxetable*}{lccccccccccc}
\tablecaption{Dense cores in W-HFS and E-hub \label{tab:ppcore}}
\tablehead{
\colhead{Core ID} &
\multicolumn{2}{c}{Position} &
\colhead{} &
\multicolumn{2}{c}{Size} & 
\colhead{PA} &
\colhead{$M$} & 
\colhead{$V_{\rm peak}$} &
\colhead{$\sigma_{\rm NT}$} &
\tabularnewline 
\cline{2-3} \cline{5-6} 
\colhead{} &
\colhead{RA} &
\colhead{Dec} &
\colhead{} &
\colhead{Major} & 
\colhead{Minor} &
\colhead{} &
\colhead{} &
\colhead{} &
\colhead{} &
\colhead{} \tabularnewline
\colhead{} &
\colhead{(hh:mm:ss)} &
\colhead{(dd:mm:ss)} & 
\colhead{} &
\colhead{(pc)} & 
\colhead{(pc)} & 
\colhead{(deg.)} &
\colhead{($M_{\odot}$)} &
\colhead{($\kms$)} &
\colhead{($\kms$)} &
\colhead{}}
\startdata
C1 & 21:44:58.5 & +47:42:02.1 & ~ & 0.05 & 0.03 & 28 & 2.4$\pm$0.6 & 1.63 & 0.24 \\ 
C2 & 21:44:52.3 & +47:40:28.8 & ~ & 0.06 & 0.04 & 96 & 8.8$\pm$2.2 & 4.28 & 0.42 \\ 
C3 & 21:44:59.1 & +47:39:50.2 & ~ & 0.08 & 0.04 & 115 & 7.5$\pm$1.9 & 4.12 &0.40 \\ 
C4 & 21:44:55.3 & +47:39:19.8 & ~ & 0.04 & 0.03 & 27 & 2.9$\pm$0.7 & 4.38 & 0.09 \\ 
N-filament$^{\ast}$ & 21:45:02.1 & +47:43:02.6 & ~ & 1.04 & 0.05 & 39 & 5.3$\pm$1.2 & 1.60 & 0.23 & \\ 
W-hub & 21:44:55.6 & +47:40:01.3 & ~ & 0.14 & 0.06 & 123 & 19.6$\pm$3.9 & 4.20 & 0.43 \\ 
E-47$^{\dagger}$ & 21:47:22.8 & +47:32:12.0 & ~ & 0.11 & 0.10 & 86 & 46$\pm$11 & -- & 0.36 \\
E-hub$^{\dagger\dagger}$ & 21:47:22.7 & +47:32:18.0 & ~ & 0.14 & 0.20 & 62 & 59$\pm$11 & -- & 0.36 \\ 
\enddata
\tablecomments{
The given sizes of major and minor axes are deconvolved with the beam, i.e., $size = \sqrt{size_{\rm uncorrected}^{2} - beam^{2}}$ \citep{berry2015}. The uncertainty of mass is assigned by propagating the errors of observational data, distance, and dust temperature. The peak velocity ($V_{\rm peak}$) and nonthermal velocity dispersion ($\sigma_{\rm NT}$) are derived with the $\ceo~(3-2)$ data. See Section~\ref{ssec:coreid} for more details. $^{\ast}$~The major and minor sizes of N-filament are the length and width estimated with {\sc filfinder} algorithm. $^{\dagger}$~E-47 was identified as the Clump Number 47 in the E-hub by \citet{johnstone2017} (see Section~\ref{ssec:compew}). The size and mass of E-47 are re-estimated with distance of 600~pc, and the beam correction is done for the size. $^{\dagger\dagger}$~E-hub is the region including clumps 45, 46, 47, 48, 52, and 53 whose clump numbers are given by \citet{johnstone2017}, and its major and minor sizes are obtained from \textsc{FellWalker} algorithm. The mass of E-hub is re-estimated from the sum of the clumps' masses with the distance of 600~pc.}

\end{deluxetable*}

\begin{figure*} \epsscale{1.17}
\plotone{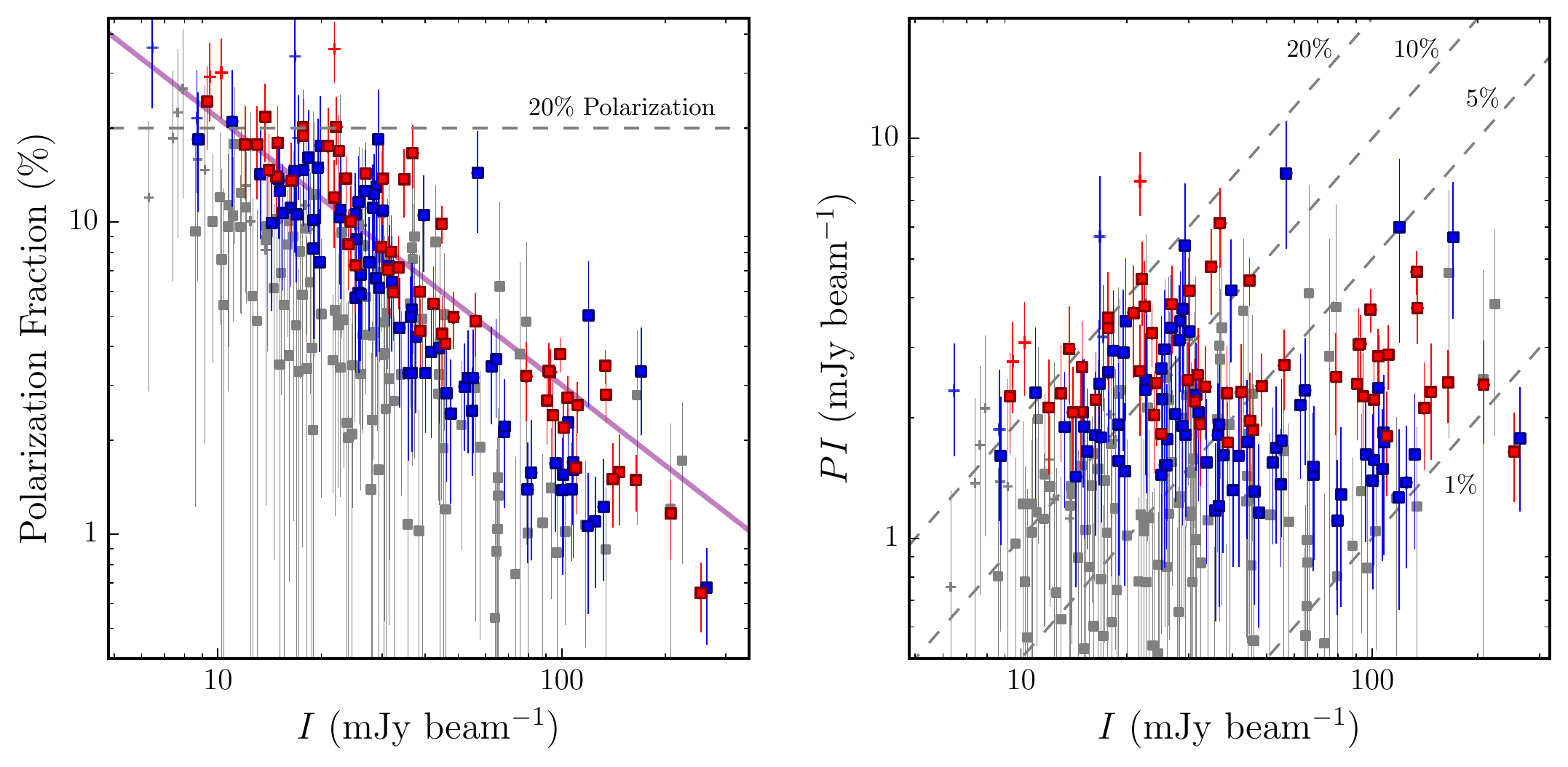}
\caption{{\it Left:} Relationship between the debiased polarization fraction and the Stokes {\em I} intensity. The gray, blue, and red colors indicate the polarization vectors with $P / \sigma_{P} \leq 2$, $2 < P / \sigma_{P} \leq 3$, and $P / \sigma_{P} > 3$, respectively. The crosses and squares are for the vectors with $5 < I / \sigma_{I} \leq 10$ and $I / \sigma_{I} > 10$, respectively. The solid line indicates the power-law fit between the polarization fraction $P$ and the Stokes $I$ total intensity of the polarization vectors with $P / \sigma_{P} > 2$ and $I / \sigma_{I} > 10$ ($P \propto I^{-0.86}$). {\it Right:} Relationship between the debiased polarized intensity ($PI$) and the Stokes {\em I} intensity. The polarization fractions from 1\% to 20\% are given with dashed lines.  
\label{fig:i_pi_lsq}}
\end{figure*}

The 850~$\mu$m emission appears to trace the dense cores on the filaments. The W-hub has the dust emission of $\sim 10 - 400$~mJy~beam$^{-1}$, which is similar to that of the E-hub shown in W2019. We identified dense cores by applying \textsc{FellWalker} clump-finding algorithm \citep{berry2015} to the 850~$\mu$m emission. Pixels with intensities $> 3 \sigma$ are used to find cores, and an object having a peak intensity higher than 10$\sigma$ and a size larger than 2$\times$beam size of 14.1$^{\as}$ is identified as a real dense core. In the case that there are neighboring peaks, these two peaks are considered to separately exist if the difference between the peak values and the minimum value (dip value) between the peaks is larger than 2$\sigma$. We found a few tens of dense cores in the circle of 11$^\prime$ diameter and 3 dense cores in the W-hub. In this paper, we analyzed the central four dense cores named as C1, C2, C3, and C4 from north to south.

C1 and C2 are found to contain Young Stellar Objects (YSOs) from Spitzer data \citep{harvey2008}, while C3 and C4 are more likely starless. C1 has one Class~I YSO while C2 has multiple YSOs, two Class~I and one Class~II YSOs. Particularly one YSO in C2, IRAS 21429+4726, shows a prominent outflow \citep{dobashi2001}, suggesting that C2 is a more active star-forming region than C1. The blue- and red-shifted lobes driven by IRAS 21429+4726 are indicated with two colored arrows in Figure~\ref{fig:850cores}.

We calculated the mass of 850~$\mu$m core with a following equation \citep[e.g.,][]{hildebrand1983}:
\begin{equation}
{\it M} = \frac{S_{\nu} ~d^{2}}{\kappa_{\nu}~ B_{\nu}(T_{\rm d})}, \label{eq:m850}
\end{equation}
where $S_{\nu}$, $\kappa_{\nu}$, $B_{\nu}$, $T_{\rm d}$, and $d$ are the integrated flux density, opacity, Planck function at the wavelength of 850~$\mu$m, the dust temperature, and the distance, respectively. The dust opacity is obtained by the equation of $\kappa_{\nu} = 0.1(\nu / 10^{12} \rm Hz)^{\beta} \rm cm^{2}~g^{-1}$ assuming a dust-to-gas ratio of 1:100 \citep{beckwith1991}, and the dust emissivity index of $\beta=2$ \citep{draine1984}. The dust temperature was taken from Herschel data \citep{andre2010,arzoumanian2011} after convolution with the JCMT resolution of 14.1$^{\as}$ using the Gaussian convolution kernel. The masses of the cores range between $\sim$2 and 9~$M_{\odot}$. 

The W-hub and N-filament are also identified with {\sc FellWalker} algorithm. Pixels with intensity $> 0.5 \sigma$ are used, and the resulting coverages of the W-hub and N-filament are presented with white polygons in Figure~\ref{fig:850cores}. It shows that W-hub covers the area of C2, C3, and C4, and N-filament includes three cores, i.e., C1, another core to the north (N-C1), and the other one to the west of C1 (W-C1). Their masses are calculated using Equation~\ref{eq:m850}. The length and width of N-filament are derived with {\sc filfinder} algorithm \citep{koch2015}.

Figure~\ref{af:c18oprof} shows the $\ceo~(3-2)$ integrated intensity map and the averaged spectra of the N-filament, W-hub, and cores. The moment~0 map is integrated over the velocity range from 0.5 to 6.0~$\kms$. As shown in the Figure, 
the spectrum of C3 looks like having a single Gaussian component, and its peak velocity (4.12~$\kms$) is well consistent to the core velocity derived with $\nthp~(1-0)$ \citep[4.20~$\kms$;][]{chung2021}. But, the spectra of some cores have the other secondary velocity component. It is revealed that some filaments have multiple velocity substructures of fibers \citep[see][and references therein]{pineda2022}, and the $\ceo~(1-0)$ observations of this region found that there are overlaps of different velocity components in the plane of the sky \citep{chung2021}. Hence, to estimate the $\ceo~(3-2)$ line width associated with W-hub, N-filament, and cores, we performed a multicomponent Gaussian fit to the averaged spectra extracted over the area of W-hub, N-filament, and each core.

The fitting results are presented with red and blue lines on the observed spectra in the Figure~\ref{af:c18oprof}. We compared the resulting peak velocities of Gaussian decomposed components with the velocities of core materials derived with the $\nthp~(1-0)$ line data \citep{chung2021}. The major components of C1 (1.63~$\kms$), C2 (4.28~$\kms$), and C4 (4.38~$\kms$) agree to the $\nthp~(1-0)$ peak velocity of the cores (1.70, 4.20, 4.20, and 4.20~$\kms$, respectively). Hence, we used the line width of major component as a representative of gas velocity dispersion of the three cores. 
The velocity dispersions ($\sigma_{\rm obs}$) are 0.25$\pm$0.09, 0.42$\pm$0.10, 0.40$\pm$0.10, and 0.11$\pm$0.08~$\kms$ for C1, C2, C3, and C4, respectively. The major components of W-hub (4.24~$\kms$) and N-filament (1.60~$\kms$) are well matched to those of the cores included in them, and $\sigma_{\rm obs}$ of the W-hub and N-filament are 0.34$\pm$0.12 and 0.23$\pm$0.08~$\kms$, respectively. The given uncertainty of the velocity dispersion is the standard deviation of the velocity dispersions of the spectra in each core.

The total 1-dimensional velocity dispersion ($\sigma_{\rm tot}$) is given with the sum of non-thermal ($\sigma_{\rm NT}$) and thermal ($\sigma_{\rm T}$) components in quadrature of $\sigma_{\rm tot}^{2} = \sigma_{\rm NT}^{2} + \sigma_{\rm T}^{2}$ \citep{myers1983}. The thermal velocity dispersion of the observed molecule is 
\begin{equation}
\sigma_{\rm T, obs} = \sqrt{\frac{k_{\rm B} T}{\mu_{\rm obs} m_{\rm H}}},
\end{equation}
where $k_{\rm B}$, $T$, $\mu_{\rm obs}$, and $m_{\rm H}$ are the Boltzmann constant, the gas temperature, the atomic weight of the observed molecule, and the hydrogen mass, respectively. Then, the non-thermal velocity dispersion can be calculated by extracting the thermal velocity dispersion from the observed total velocity dispersion:
\begin{equation}
\sigma_{\rm NT} = \sqrt{\sigma_{\rm obs}^{2} - \frac{k_{\rm B} T}{\mu_{\rm obs} m_{\rm H}}}, \label{eq:sigmaNT}
\end{equation}
where $\sigma_{\rm obs}$ is the observed velocity dispersion from the line width of the observed spectrum ($\sigma_{\rm obs} = \Delta v / \sqrt{8 \rm ln 2}$). We used $\sigma_{\rm obs}$ for the $\ceo~(3-2)$ line. 
The estimated properties of the W-hub, N-filament, and cores are listed in Table~\ref{tab:ppcore}. \\

\subsection{Polarization Properties} \label{ssec:pp}

\begin{figure*} \epsscale{1.17}
\plotone{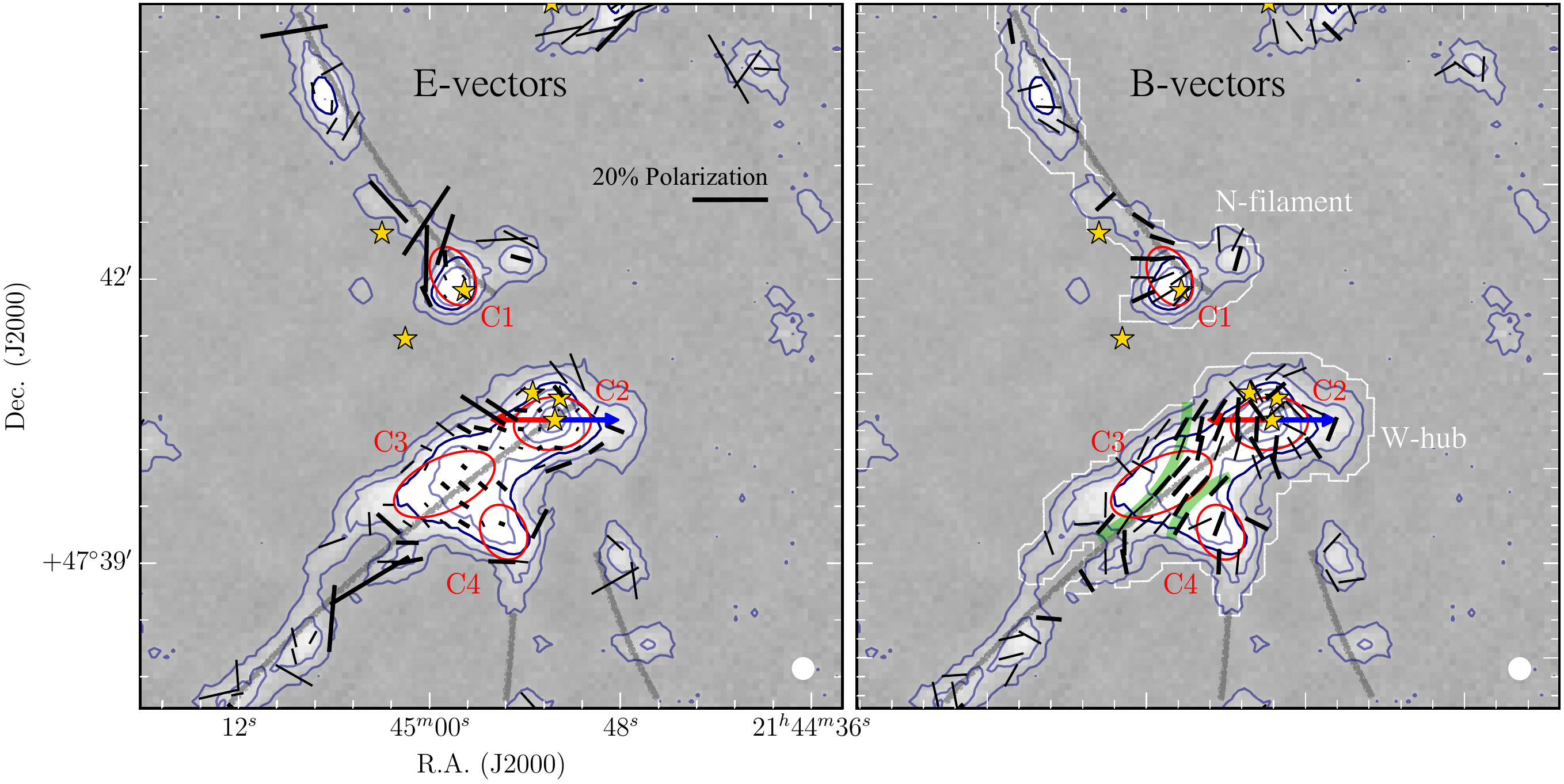}
\caption{Polarization vectors (left) and magnetic field vectors (right) on 850 $\mu$m images. The contour levels are 3, 10, 20, 30, 60, and 90$\times \sigma$ (1$\sigma=3.1$~mJy~beam$^{-1}$). The polarization vectors are chosen to have $I / \sigma_{I} > 10$ threshold. The thin and thick lines denote the vectors with $2 < P / \sigma_{P} \leq 3$ and $P / \sigma_{P} > 3$, respectively. The magnetic field orientations are assumed to be perpendicular to the observed polarization and their lengths are equally given to better show the magnetic field orientation. The red ellipses, yellow stars, red- and blue-arrows, and gray lines indicate the 850~$\mu$m cores, YSOs, outflow, and the guide lines for the $\ceo$ filaments as shown in Figure~\ref{fig:850cores}, respectively. The green lines in the right panel are drawn by eye to show the curved shape of the magnetic field. The JCMT beam (14.1$^{\as}$) is presented on the bottom right corner of the figure. \label{fig:pvectors}}
\end{figure*}

Figure~\ref{fig:i_pi_lsq} shows the polarization fraction ($P$) as a function of the total intensity ($I$). The dependence of $P$ on $I$ is described with a power-law index $\alpha$ of $P \propto I^{- \alpha}$. $\alpha$ is closely related to the grain alignment efficiency. If the dust grains align 
in a same fashion at all optical depths, $\alpha$ would equal zero, but if the grain alignment linearly decreases along with the increasing optical depth, $\alpha$ would be 0.5. The unity of $\alpha$ indicates that the grains align only in the thin layer at the surface of the cloud, while grains at higher densities do not align in any special direction. In the W-HFS of IC~5146, it shows a decreasing trend of polarization fraction at higher intensity regions ($\alpha=0.86$), suggesting higher degree of depolarization at the higher density region. The depolarization at the denser region is also well represented in the relationship between the polarized intensity and the total intensity shown in the right panel of Figure~\ref{fig:i_pi_lsq}. The polarization fractions at $I \lesssim 40 \rm ~mJy~beam^{-1}$ are in a range of 5 and 20\%, while those at $I > 40 \rm ~mJy~beam^{-1}$ are less than 5\%. 

The polarization vectors are presented in the left panel of Figure~\ref{fig:pvectors} on the 850~$\mu$m image. Polarized emission is found at the less dense filaments as well as at the more dense hub. As shown in Figure~\ref{fig:i_pi_lsq}, Figure~\ref{fig:pvectors} also displays that the polarization fraction decreases at the dense core regions. \\

\subsection{Magnetic Field Morphology and Strength}

\subsubsection{Magnetic Field Morphology} \label{sss:bmor}

Magnetic field orientations can be obtained by rotating submillimeter polarization vectors by 90 degree. The right panel of Figure~\ref{fig:pvectors} shows the magnetic field vectors at the W-HFS. Around C1, the main direction of magnetic field is southeast-northwest. This is nearly perpendicular to the filament direction of northeast-southwest. On the contrary, the magnetic field morphology around the cores in the hub is much more complex. The main direction is likely to be south-north, but the magnetic field vectors with east-west direction can be found too. There are two main characters of magnetic fields in the W-hub. One is the abrupt changes of the magnetic field vectors around C2, and the other is the curved magnetic field in the near vicinity of C3. The sudden change of the orientations of magnetic field vectors near the center of C2 is seemingly related to the bipolar outflows observed in CO \citep{dobashi2001}. The outflow is along the east-west direction. The magnetic field vectors show a slightly curved, hourglass morphology at the northern and western region of C2. Hence, the magnetic fields lines near C2 are possibly getting modified by the outflows. 

The magnetic fields near C3 seem to have a pinched shape as presented with the green lines in Figure~\ref{fig:pvectors}. This hourglass shape of magnetic field morphology can be the result of gravitational contraction of C3 \citep[e.g.,][]{pattle2017,wang2019}. Another possibility is that the active gas motions such as infalls and accretion flows which are observed in the W-hub modify the magnetic field. Infall signatures are observed around the cores of C3 and C4 in the W-hub by the $\hcop~(1-0)$ molecular line observations, and the velocity gradients of $\ceo~(1-0)$ in the W-HFS are found, implying the existence of possible accretion flows from filaments to the hub \citep[][]{chung2021}. Numerous observations propose that the infall motion and accretion flow can modify the magnetic field \citep[e.g.,][]{pillai2020}. The curved magnetic field lines going with the elongated filamentary structure in the W-hub can be an evidence of the modification of the magnetic fields due to the gas motions of infall and accretion flows. \\ 

\subsubsection{Magnetic Field Strength}

We measured the magnetic field strengths of the cores C1, C2, and C3 using the David-Chandrasekhar-Fermi (DCF) method \citep{davis1951,chandrasekhar1953}. The magnetic field strength of C4 is not calculated since the number of magnetic field vectors are too small. The total magnetic fields of the W-hub and N-filament are also estimated. By assuming that the underlying magnetic field is uniform but distorted by the turbulence, the DCF method estimates the magnetic field strength in the plane of the sky ($B_{\rm pos}$) in $\mu$G from the magnetic field angular dispersion ($\delta \phi$), the velocity dispersion of the gas ($\sigma$), and the gas density ($\rho$) using the equation 
\citep{crutcher2004etal}:
\begin{align}
	B_{\rm pos} &= Q_{\rm c} \sqrt{4 \pi \rho} \frac{\sigma}{\delta \phi} \nonumber  \\
	&\approx 9.3 \sqrt{\bar{n}_{\rm H_{2}}} \frac{\Delta v}{\delta \phi} ,
\end{align}
\noindent where $Q_{\rm c}$ is the correction factor for the underestimation of angular dispersion in polarization map due to the beam integration effect and hence overestimation of the magnetic field strength, adopted as 0.5 from \citet{ostriker2001}. $\bar{n}_{\rm H_{2}}$ is the mean volume density of the molecular hydrogen in cm$^{-3}$, and $\Delta v = \sigma_{\rm NT} \sqrt{\rm 8 ln 2}$ in $\kms$. 

\begin{figure*} \epsscale{1.17}
\plotone{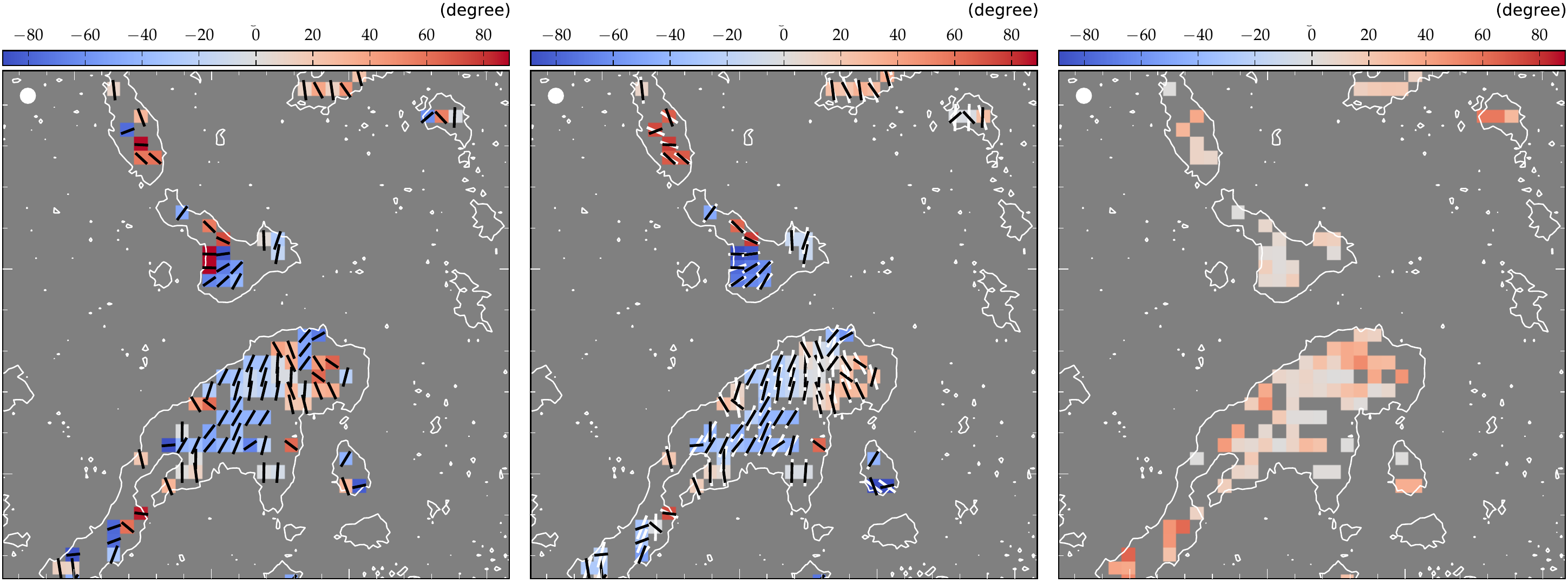}
\caption{The position angles of magnetic field measured clockwise with respect to the north are shown with color maps. The left and the middle panels show the position angles of observed magnetic field and those of smoothed magnetic field with a 3$\times$3 pixel boxcar filter. The right panel presents the residual map by subtracting the smoothed map from the observed map. The observed B-field vector (black in the left and the middle) and the smoothed B-field vector (white in the middle) are presented on the images. \label{fig:bgsubtract}}
\end{figure*}

We applied an unsharp-masking method to remove the underlying ordered magnetic field structure, and then measured the magnetic field angle dispersion \citep{pattle2017}. Firstly, we estimated the large scale, background magnetic field structure by smoothing the magnetic field map with a 3$\times$3 pixel boxcar filter ($36^{\as} \times 36^{\as}$)\footnote{The size of boxcar filter is chosen to remove the underlying curved magnetic fields that seem to be caused by the gas motions such as outflow, infall, and accretion flows as mentioned in Section~\ref{sss:bmor}. 
\citet{pattle2017} reported that the use of larger filter sizes than 3$\times$3 pixel can cause an overestimation of angular dispersion for even shallow field curvature. When applying 5$\times$5 pixel boxcar filter to our data, we failed to reproduce the curved magnetic field shapes in the core regions, especially around C2, and resulting angular dispersions are found to be larger than those measured from 3$\times$3 pixel boxcar filter.}. Then, the observed magnetic field map was subtracted from the smoothed map. Finally, the angular dispersion is measured from the residual map. The standard deviation of the polarization angle error is given as the uncertainty of the estimated angular dispersion in Table~\ref{tab:bfield_hf}. The angular dispersions of the regions range from $\sim$7 to 20 degree and thus the DCF method is found to be well applicable. 

The mean H$_{2}$ volume densities ($\bar{n}_{\rm H_{2}}$) of cores and W-hub are estimated with the total mass and ellipsoid volume assuming that the thickness 
is equal to the geometric mean of the observed major and minor axis obtained from the 2D Gaussian fit. The volume of assumed spheroid is same to that of sphere having a radius of geometric mean of the observed semi-major and semi-minor size, but larger and smaller than those of oblate and prolate spheroids, respectively. The mean of volume difference of the assumed spheroid to the oblate and prolate spheroids is used for the propagation error of $\bar{n}_{\rm H_{2}}$. $\bar{n}_{\rm H_{2}}$ of the N-filament is estimated with the total mass and cylindrical volume assuming that the radius is same to the half of the filament's width. $\Delta v$ was estimated from the non-thermal velocity dispersion obtained in the Equation~\ref{eq:sigmaNT} using the measurement of the line width of the averaged spectrum of $\ceo~(3-2)$ line. 

The applied $\bar{n}_{\rm H_{2}}$ and $\Delta v$ and the measured magnetic field strengths are tabulated in Table~\ref{tab:bfield_hf}. The measured magnetic field strengths of the N-filament and W-hub are 80 and 600~$\mu$G, respectively, and those of the cores in the W-HFS are found to range between 0.8 and 1.2~mG which is close to the median value of the magnetic field strengths of molecular clouds belonging to the Gould Belt studied by JCMT BISTRO survey (120$\pm 60 \mu$G in the Perseus B1 by \citet{coude2019}; 6.6$\pm 4.7$~mG in the OMC~1 by \citet{pattle2017}). \\

\begin{deluxetable*}{lcccccccc} \label{tab:bfield_hf}
\tablecaption{Magnetic Field Strengths and Energies \label{tab:bfield}}
\tablehead{
\colhead{} &
\colhead{N-filament} &
\colhead{W-hub} &
\colhead{E-hub$^{\dagger}$} &
\colhead{C1} &
\colhead{C2} &
\colhead{C3} &
\colhead{E-47$^\dagger$} 
}
\startdata
Number of B-vectors & 57 & 20 & 45 & 9 & 25 & 21 & 45 \\
$\delta \phi$ (degree) & 13$\pm$2 & 16$\pm$3 & 17.4$\pm$0.6 & 6.5$\pm$0.9 & 20$\pm$3 & 13$\pm$2 & 17.4$\pm$0.6 \\ 
$\Delta v$ ($\kms$) & 0.5$\pm$0.2 & 1.0$\pm$0.3 & 0.9$\pm$0.2 & 0.6$\pm$0.2 & 1.0$\pm$0.2 & 0.9$\pm$0.2 & 0.9$\pm$0.2 \\ 
$\bar n_{\rm H_{2}}$ ($\rm 10^{5} ~ cm^{-3}$) & 0.5$\pm$0.1 & 10$\pm$5 & 5$\pm$1 & 22$\pm$8 & 34$\pm$12 & 17$\pm$8 & 14$\pm$3 \\ 
\hline 
$B_{\rm pos}$ ($\mu$G) & 80$\pm$30 & 600$\pm$200 & 320$\pm$60 & 1200$\pm$600 & 800$\pm$300 & 900$\pm$300 & 500$\pm$100 \\ 
$\lambda$  & 0.2$\pm$0.1 & 1.2$\pm$0.9 & 1.9$\pm$0.7 & 0.5$\pm$0.3 & 1.5$\pm$0.8 & 0.8$\pm$0.6 & 2.2$\pm$0.7 \\ 
$V_{\rm A}$ ($\kms$) & 0.6$\pm$0.3 & 1.0$\pm$0.4 & 0.8$\pm$0.1 & 1.4$\pm$0.6 & 0.8$\pm$0.3 & 1.1$\pm$0.4 & 0.8$\pm$0.1 \\ 
$M_{\rm A}$ & 0.4$\pm$0.2 & 0.4$\pm$0.2 & 0.5$\pm$0.2 & 0.2$\pm$0.1 & 0.5$\pm$0.2 & 0.3$\pm$0.2 & 0.5$\pm$0.2 \\ 
\hline 
$E_{\rm G}$ (M$_{\odot}$ km$^{2}$ s$^{-2}$) & 0.12$\pm$0.05 & 22$\pm$9 & 100$\pm$60 & 0.9$\pm$0.5 & 9$\pm$4 & 5$\pm$3 & 100$\pm$50 & \\ 
$E_{\rm K}$ (M$_{\odot}$ km$^{2}$ s$^{-2}$) & 0.5$\pm$0.3 & 7$\pm$4 & 16$\pm$7 & 0.4$\pm$0.2 & 3$\pm$1 & 2$\pm$1 & 13$\pm$6 & \\ 
$E_{\rm M}$ (M$_{\odot}$ km$^{2}$ s$^{-2}$) & 1.1$\pm$0.5 & 9$\pm$4 & 18$\pm$5 & 2$\pm$1 & 3$\pm$1 & 5$\pm$2 & 14$\pm$4 \\ 
\hline 
$|(2E_{\rm K}+E_{\rm B})/E_{\rm G}|$ & 18$\pm$11 & 1.0$\pm$0.6 & 0.5$\pm$0.3 & 3$\pm$2 & 0.9$\pm$0.6 & 2$\pm$1 & 0.4$\pm$0.2 & \\ 
\enddata
\tablecomments{$^\dagger$The magnetic field strengths of E-hub and E-47 are re-estimated with $\delta \phi$ and $\sigma_{\rm obs}$ given in \citet{wang2019}. See Section~\ref{ssec:compew} for details.}
\end{deluxetable*}
\vspace{5mm}


\section{Analysis} \label{sec:anal}

\subsection{Magnetic field strength versus Gravity} \label{sec:bvsg}

The observed mass-to-magnetic flux ratio is compared with the critical mass-to-magnetic flux ratio to discuss the relative importance of magnetic fields and gravity. 
The observed mass-to-magnetic flux ratio in units of the critical ratio ($\lambda_{\rm obs}$) is described as follows \citep{crutcher2004etal}:
\begin{equation}
	\lambda_{\rm obs} = \frac{(M/\Phi)_{\rm obs}}{(M/\Phi)_{\rm crit}}.
\end{equation}
The observed mass-to-magnetic flux ratio is 
\begin{equation}
	(M/\Phi)_{\rm obs} = \frac{\mu m_{\rm H} N_{\rm H_{2}}}{B_{\rm pos}},
\end{equation}
where $\mu$ is the mean molecular weight of 2.8 and $N_{\rm H_{2}}$ is the H$_{2}$ column density, and the critical mass-to-magnetic flux ratio is 
\begin{equation}
	(M/\Phi)_{\rm crit} = \frac{1}{2 \pi \sqrt{G}}.
\end{equation}

\noindent \citet{crutcher2004} proposed $\lambda_{\rm obs} = 7.6 \times 10^{-21} N_{\rm H_{2}} / B_{\rm pos}$ with $N_{\rm H_{2}}$ in cm$^{-2}$ and $B_{\rm pos}$ in $\mu$G. The real mass-to-magnetic flux ratio can be estimated using a statistical mean correction factor of one third accounting for the random inclinations for an oblate spheroid core, flattened perpendicular to the orientation of the magnetic field \citep[$\lambda = \lambda_{\rm obs} / 3$;][]{crutcher2004etal}. 

The estimated mass-to-magnetic flux ratios are given in Table~\ref{tab:bfield_hf}. $\lambda$ of N-filament and W-hub are 0.2$\pm$0.1 and 1.2$\pm$0.9, and they are magnetically subcritical and supercritical, respectively. $\lambda$ of C1, C2, and C3 are 0.5$\pm$0.3, 1.5$\pm$0.8, and 0.8$\pm$0.6, respectively, implying that C1 is magnetically subcritical while C2 and C3 are marginally supercritical within the uncertainty. The magnetic field strength measured from the DCF method tends to be overestimated due to the limited angular resolution and the possible smoothing effect along the line of sight which cause underestimation of angular dispersion \citep[e.g.,][]{heitsch2001,ostriker2001,crutcher2012}. Meanwhile, the non-thermal velocity dispersion used in the estimation of magnetic field strength is possibly overestimated because motions such as mass flow, infall, and outflow are possibly added into the turbulence motion. Hence, the magnetic field strengths of the cores presented here would be the upper limit of the true value and the mass-to-magnetic flux ratio would be the lower limit. \\

\subsection{Magnetic field strength versus Turbulence} \label{sec:bvst}

The importance of magnetic fields with respect to the kinetic energy is investigated using the Alfv\'enic Mach number ($M_{\rm A}$) with the following equation:
\begin{equation}
	M_{\rm A} = \frac{\sigma_{\rm NT}}{V_{\rm A}},
\end{equation}
where $\sigma_{\rm NT}$ is the non-thermal velocity dispersion and $V_{\rm A}$ is the Alfv\'en velocity. Alfv\'en velocity is estimated by 
\begin{equation}
	V_{\rm A} = \frac{B}{\sqrt{4 \pi \bar{\rho}}} 
\end{equation}
where $\bar{\rho}$ is the mean density ($\mu m_{\rm H} \bar{n}_{\rm H_{2}}$). For the total magnetic field strength of $B$, the statistical average value of $B_{\rm pos}$, $(4/\pi)B_{\rm pos}$, is used \citep{crutcher2004etal}. The calculated Alfv\'en velocities and Mach numbers are presented in Table~\ref{tab:bfield_hf}. $M_{\rm A}$ of the N-filament, W-hub, and cores are found to range between 0.2 to 0.5, being sub-Alv\'enic. Therefore, the magnetic field dominates turbulence in W-HFS. \\ 

\subsection{Energy balance} \label{ssec:energies}

We estimated the gravitational, kinematic, and magnetic energies of the N-filament, W-hub, and cores in the W-HFS 
following \citet{mckee2007}. The virial theorem to discuss the relative importance of gravitational, kinematic, and magnetic energies in a molecular cloud can be written as
\begin{equation}
	\frac{1}{2} \ddot{I} = 2E_{\rm K} + E_{\rm B} + E_{\rm G} ,
\end{equation}
where $E_{\rm K}$, $E_{\rm B}$, and $E_{\rm G}$ are the total kinematic energy, magnetic energy, and gravitational potential energy, respectively. The quantity $I$ is proportional to the inertia tensor of the cloud, and hence the positive and negative $\ddot{I}$ present the acceleration of expansion and contraction of the cloud. 

The total kinematic energies of the W-hub and cores are estimated as
\begin{equation}
	E_{\rm K}^{\rm sphere} = \frac{3}{2} M \sigma_{\rm tot}^{2} ,
\end{equation}
and that of the N-filament is derived as
\begin{equation}
	E_{\rm K}^{\rm cylinder} = M \sigma_{\rm tot}^{2} ,
\end{equation}
where $\sigma_{\rm tot}$ is the observed total velocity dispersion \citep[e.g.][]{fiege2000i}. The observed total velocity dispersion is estimated with the mean free particle of molecular weight $\mu_{\rm p}$=2.37 \citep{kauffmann2008} by the equation:
\begin{equation}
	\sigma_{\rm tot} = \sqrt{\sigma_{\rm NT}^{2} + \frac{k_{\rm B} T}{\mu_{\rm p} m_{\rm H}}} .
\end{equation}

The magnetic energy is calculated with the equation of
\begin{equation}
	E_{\rm B} = \frac{1}{2} M V_{\rm A}^{2}. 
\end{equation}

The gravitational energies for the W-hub and cores are estimated from the equation of
\begin{equation}
	E_{\rm G}^{\rm sphere} = - \frac{3}{5} \frac{GM^{2}}{R}. \label{eq:eqegs}
\end{equation}
\noindent The geometric mean values of the semi-major and semi-minor sizes of the W-hub and cores are used for $R$. The gravitational energy for the N-filament is calculated from the equation of
\begin{equation}
	E_{\rm G}^{\rm cylinder} = - \frac{GM^{2}}{L} \label{eq:eqegf},
\end{equation}
where $L$ is the length of filament \citep{fiege2000i}.

In these calculations of energies, the surface kinetic energy is ignored, and thus the estimated total kinematic energy should be treated only in the aspect of self-stability in the enclosed region \citep{wang2020b}. As the sign of $\ddot{I}$ indicates the expansion (plus) or contraction (minus) of the core, thus $|(2E_{\rm K}+E_{\rm B})/E_{\rm G}|$ can be used as an indicator of the expansion ($> 1$) or contraction of the core ($< 1$). The estimated quantities are tabulated in Table~\ref{tab:bfield_hf}. \\

\section{Discussion} \label{sec:disc}

\subsection{HFSs in the both ends of Streamer} \label{ssec:compew}

\begin{sidewaysfigure*}
\begin{center}
\vspace{-3.3in}
\includegraphics[width=\textwidth,keepaspectratio]{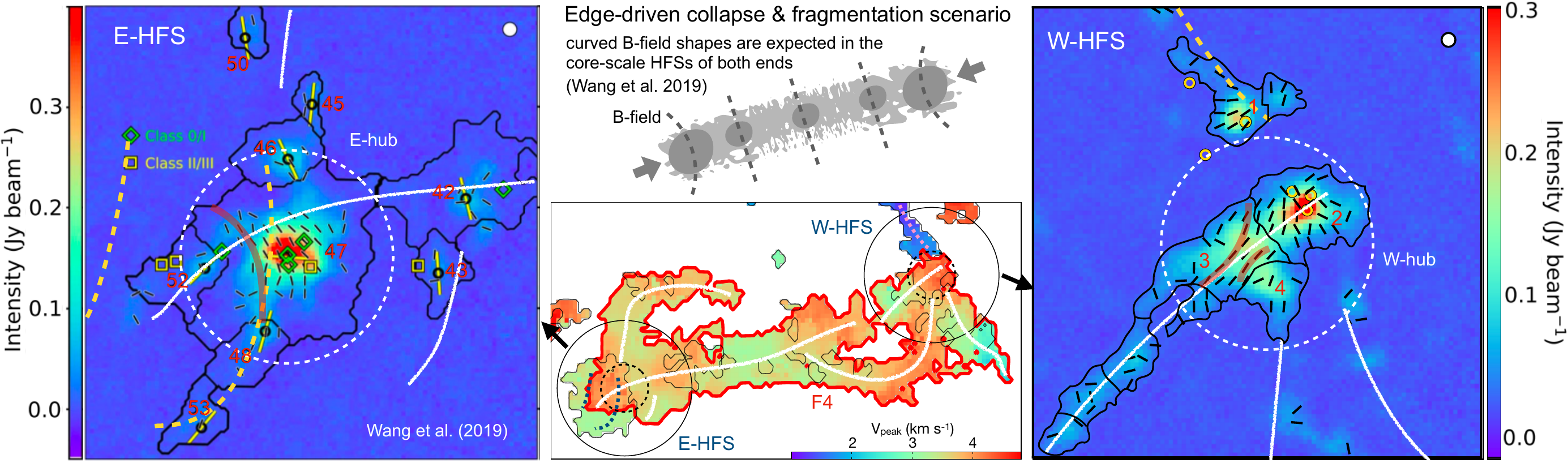}
\end{center}
\caption{The eastern-HFS (E-HFS) and western-HFS (W-HFS) regions of the IC~5146 Streamer which have been observed with JCMT/POL-2. The hub regions (E- and W-hubs) are indicated with white dashed circles with a 3$^{\prime}$ diameter on the E- and W-HFSs' 850~$\mu$m images of the left and right panels. The white circles at the top right corner of the left and right panels show the POL-2 850~$\mu$m beam size of 14.1~arcsec. The white solid curves are the guide lines of velocity coherent filament F4 and the dashed curves are those of filaments having different velocities to that of F4. The red lines in the left and right panels are given to show the curved magnetic fields in the hubs. {\it Left}: The magnetic field orientation map of the E-HFS of IC~5146 taken from W2019. {\it Middle}: The schematic figure of edge-driven collapse and fragmentation scenario from W2019 and $\ceo~(1-0)$ peak velocity map of the Streamer by the TRAO FUNS \citep{chung2021}. The velocity coherent filament, F4, where the E- and W-hubs are located is outlined with red. The filaments and clumps with different velocities from that of F4 are overlapped in the line-of-sight direction, and their velocity fields and outlines are drawn with images on the background and thin solid lines in the foreground of F4, respectively). The two navy circles indicate the E- and W-hub regions observed with JCMT/POL-2. The smaller dashed circles represent the inner 3$^{\prime}$ region with the best sensitivity. {\it Right}: The magnetic field orientation map of the W-HFS (this study).  \label{fig:ewhubs}}	
\end{sidewaysfigure*}

The dark Streamer of IC~5146 has two prominent hubs, one in the east-end and the other in the west-end. The column density map from Herschel data shows that the E-hub has the higher column density than the W-hub \citep{andre2010,arzoumanian2011}. The E- and W-hubs contain ten and four YSO candidates, respectively \citep{harvey2008,poglitsch2010,chung2021}. \citet{chung2021} have made various molecular line observations toward IC~5146 and revealed that the two hubs, located in a velocity coherent filament F4 (see the peak velocity map of F4 in Figure~\ref{fig:ewhubs}), have similar physical properties of the mass accretion rate and the nonthermal velocity dispersion. The estimated mass accretion rates from filaments to the dense cores in the E- and W-hubs are $26\pm14$ and $35\pm11~\rm M_{\odot}~ Myr^{-1}$, respectively. The nonthermal velocity dispersions of the two hubs are about 3 times of the sound speed. Meanwhile, the 850~$\mu$m image toward the E- and W-hubs shows slightly different fragmentation features. The W-hub has two cores with similar masses to each other and less massive one core, while the E-hub has a dominant clump and minor cores. 

The E-hub of IC~5146 has been investigated by W2019 as a part of BISTRO survey \citep{wardthompson2017}. They suggested the edge-driven collapse and fragmentation scenario for the formation mechanism of the two hubs in the Streamer based on the larger aspect ratio of Streamer than 5 where the edge-driven collapsing efficiently happens and the curved magnetic field shape in the hub possibly caused by the gravitational contraction. According to the edge-driven collapse and fragmentation scenario, the magnetic field of E- and W-hubs should have `)' and `(' shapes, respectively, as presented in the middle top panel of Figure~\ref{fig:ewhubs}. In that sense, we examined the magnetic field orientations in the E- and W-hubs. 
As shown in the Figure~\ref{fig:ewhubs}, the magnetic fields in the E-hub tend to be well ordered, i.e., perpendicular to the large scale filament guided with white lines. The shape of bending magnetic field in E-hub agrees to the edge-driven collapse and fragmentation scenario. In the W-hub, however, the observed shape of the magnetic field near C3 does not exactly match to the expectation of the scenario. The magnetic field in the W-hub is shown to be curved at the vicinity of C3, but has a pinched shape, guided by the red lines in the right panel of Figure~\ref{fig:ewhubs}. It is almost parallel to the direction of filament, rather than a `('-shape being perpendicular to the filament direction. The magnetic fields in the W-hub are changing its directions, and they are also connected to the elongated filamentary structures of which feature can be seen in the E-hub, too. In Section~\ref{sss:bmor}, we proposed two possibilities of the gravitational contraction and the gas motion of accretion flows. And, the accretion rates at the E-hub from filaments to cores are revealed to be similar to those at the W-hub \citep{chung2021}. Hence, the curved magnetic fields in the E- and W-hubs may be the results of the modification by the accretion flows in addition to the gravitational contraction proposed by W2019.

As a whole, the magnetic fields in the W-hub are more complex than those in the E-hub. This is seemingly related to the more complicate conditions of W-hub. E-hub has a major clump at the center (Clump Number~47, E-47 hereafter, in the top panel of the Figure~\ref{fig:ewhubs}) with minor cores at the surroundings (Clump Number~46, 48, 52, and etc.). However, W-hub has three cores (C2, C3, and C4), out of which C2 and C3 have similar masses, implying that the magnetic fields can be locally modified by the gravitational contraction among the cores in the W-hub.

We estimated the magnetic field strength of E-47 using the angular dispersion of 17.4~degree and the velocity dispersion of 0.37~$\kms$ given in W2019 and H$_{2}$ number density re-estimated with the mass and size given in \citet{johnstone2017} using the distance of 600~pc. The B-field strength of E-hub is measured with the same $\delta \phi$ and $\Delta v$ to that of E-47 but $\bar n_{\rm H_{2}}$ calculated with the mass and size of E-hub given in Table~\ref{tab:ppcore}. The mass-to-flux ratio, Alfv\'enic Mach number, $E_{\rm G}$, $E_{\rm K}$, and $E_{\rm M}$ of E-47 and E-hub are also measured, and the results are tabulated in Table~\ref{tab:bfield_hf}. 

\begin{figure*} 
\includegraphics[width=1\textwidth,keepaspectratio]{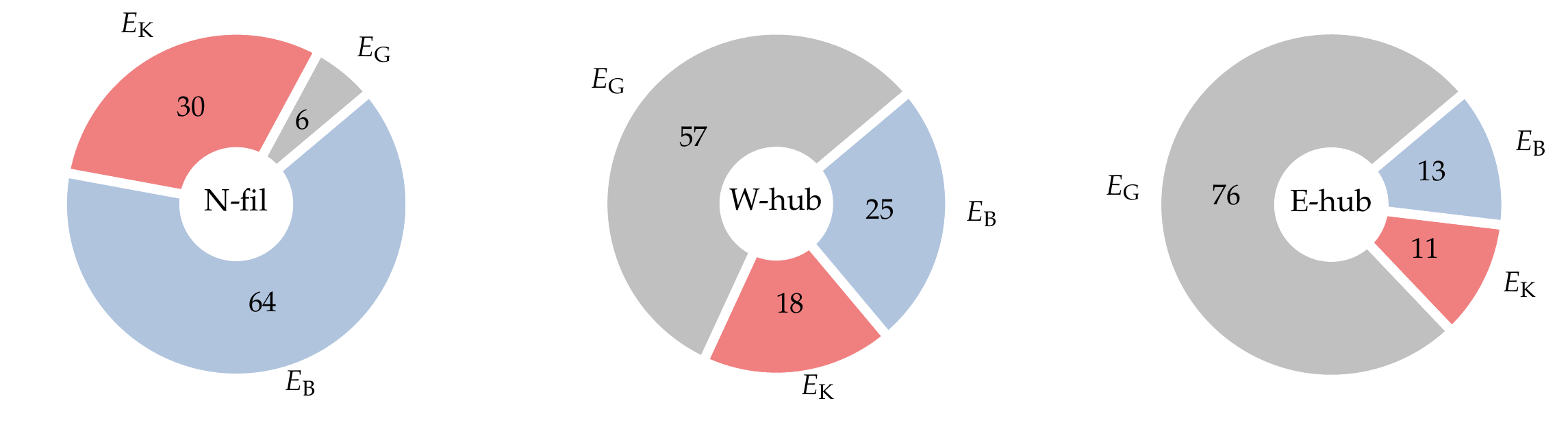} \caption{The relative importance of gravity, turbulence, and magnetic energies of N-filament, W-hub, and E-hub. The gravity energy ($E_{\rm G}$), kinematic energy ($E_{\rm K}$), and magnetic energy ($E_{\rm B}$) are presented with gray, red, and blue colors, respectively, and the relative portions are given in \%. \label{fig:energies1}}
\end{figure*}

The magnetic field strength of E-47 is $500 \pm 100$~$\mu$G which is consistent with that of W2019. The mass-to-magnetic flux ratio is $2.2 \pm 0.7$, and the Alfv\'enic Mach number is $0.5 \pm 0.2$. W2019 reported that gravity and magnetic field are currently of comparable importance in the E-hub and turbulence is less important. The mass-to-magnetic flux ratio recalculated in this study presents that E-47 is magnetically supercritical and sub-Alfv\'enic. \\

\subsection{Energy balance and fragmentation types of hubs and filaments into cores} \label{ssec:ebfrac}

This section discusses the fragmentation types of the HFSs of IC 5146 into cores with help of the energy budget. The cores would be the fragmentation results of their natal hub and filament, and their current distributions could be determined by the fragmentation types ruled by the energy balance of gravity, magnetic field, and turbulence in the hub and filament. At the same time, the cores may (or may not) fragment into smaller substructures and form protostars in the future, and their present energy balance probably gives the key to their future evolution. Here we discuss the fragmentation from filament/hub to cores in the past. As for the possibility of formation of protostars from cores in the future, we will discuss in the next section in a manner of evolution of HFSs (\S~\ref{ssec:evol}).


Recently, \citet{tang2019} proposed that the differences in relative importance of gravitational, magnetic, and turbulent energies can determine the fragmentation types of molecular clumps. They suggested three types of fragmentation, i.e., $clustered$ fragmentation where magnetic field is not so dominant that turbulence can make scattered cores, leading those to be in collapse, $aligned$ fragmentation in which magnetic field dominates turbulence and matters mainly collapse along the field lines, and $no$ fragmentation where gravity dominates both magnetic field and turbulence. 

The N-filament of W-HFS has three cores, i.e., C1, another in the north of C1 (N-C1), and the other in the west of C1 (W-C1). 
The N-filament seems to be a type of the $aligned$ fragmentation. The E- and W-hubs of IC~5146 both have multiple cores, but they show slightly different fragmentation features. As shown in Figure~\ref{fig:ewhubs}, E-hub has a dominantly large and massive core, E-47, at the center and several small cores around it. The mass of E-47 is about 10 to 100 times larger than those of the other minor cores in the E-hub. What is more interesting is that the minor cores around E-47 place at the overlapped points of the E-hub and filament having different velocity from that of E-hub. 
On the contrary to the E-hub, W-hub has three cores but no dominant core. Rather than that, two cores (C2 and C3) among the three have similar sizes and masses. Hence, E-hub and W-hub appear to be close to $no$ and $clustered$ fragmentation types, respectively.

The gravitational, kinematic, and magnetic energies are given in Table~\ref{tab:bfield_hf}. This table shows that E-hub has larger energies in gravity, kinematics, and magnetic field than those of N-filament and W-hub. This is mainly caused by the larger mass of E-hub than those of N-filament and W-hub.

Figure~\ref{fig:energies1} shows the relative distribution of $E_{\rm G}$, $E_{\rm K}$, and $E_{\rm B}$ for the regions of N-filament, W-hub, and E-hub. N-filament has the largest portion of 64\% in $E_{\rm B}$, while W- and E-hubs have the largest portion of 57\% and 76\% in $E_{\rm G}$. In case of N-filament, its energy budget and fragmentation type agree to the proposal of \citet{tang2019} where B-field dominates turbulence, local gravitational collapse happens along B-field lines, and fragments line up perpendicular to the B-field. In case of W- and E-hubs, $E_{\rm G}$ dominates both $E_{\rm K}$ and $E_{\rm B}$. However, W- and E-hubs have a mode of $clustered$ and $no$ fragmentation, respectively. For the reason why the two hubs have different fragmentation shapes even though both have dominant gravitational energy, we note that W-hub has $E_{\rm K}$ of 18\% while E-hub has only $E_{\rm K}$ of 11\%. In the proposal of \citet{tang2019}, to have $clustered$ fragmentation type, turbulence should have such meaningful importance that it can scatter the materials irregularly. Even though $E_{\rm G}$ in W-hub is relatively dominant to $E_{\rm K}$ and $E_{\rm B}$, but its portion (57\%) is smaller than that of E-hub (76\%). Moreover, the portion of $E_{\rm K}$ in W-hub is about twice larger than in E-hub. Hence, we presume that the different portions of $E_{\rm K}$ in W- and E-hubs make them have different fragmentation types.

The value of $|(2E_{\rm K}+E_{\rm B})/E_{\rm G}|$ indicates whether the interstellar clouds and/or cores are contracting ($< 1$) or expanding ($> 1$). N-filament, W-hub, and E-hub have $|(2E_{\rm K}+E_{\rm B})/E_{\rm G}| = 18 \pm 11$, 1.0$\pm$0.6, and 0.5$\pm$0.3, respectively. These values well represent the current fragmentation shapes of the filament and hubs, i.e., N-filament with $|(2E_{\rm K}+E_{\rm B})/E_{\rm G}| > 1$ dispersed and has aligned cores, while E-hub with $|(2E_{\rm K}+E_{\rm B})/E_{\rm G}| < 1$ contracted and has one massive core. W-hub has $|(2E_{\rm K}+E_{\rm B})/E_{\rm G}| \sim 1$ being on the border of contraction and dispersion and having clustered cores of similar masses. \\

\subsection{Evolution of HFSs in the Streamer} \label{ssec:evol}

Hub-filament structure is recognized as a birthplace of stars, especially high-mass stars and stellar clusters. It is suggested that the HFS forms by a layer fragmentation in which the fragmentation occurs in gas layers threaded with magnetic field \citep[e.g.,][]{myers2009}. W2019 proposed a scenario for a core-scale HFS formation in the Streamer of IC~5146 according to which the E- and W-hubs form first by edge driven fragmentation in a long filament with strong magnetic field and then further fragmentation occurs in the dense hubs, making the local magnetic field morphology to be modified. However, the magnetic fields orientations of the W-hub are found to have slightly different bending direction to that the scenario expects (see \S~\ref{ssec:compew}). Meanwhile, cloud-cloud collision is also suggested to be a possible HFS formation mechanism \citep[e.g.,][]{kumar2020}, and \citet{chung2021} suggested that the different nonthermal velocity dispersions between the hub and dense cores in them is an evidence of E- and W-HFSs formation by collision of turbulent flows.

\begin{figure*} 
\includegraphics[width=1\textwidth,keepaspectratio]{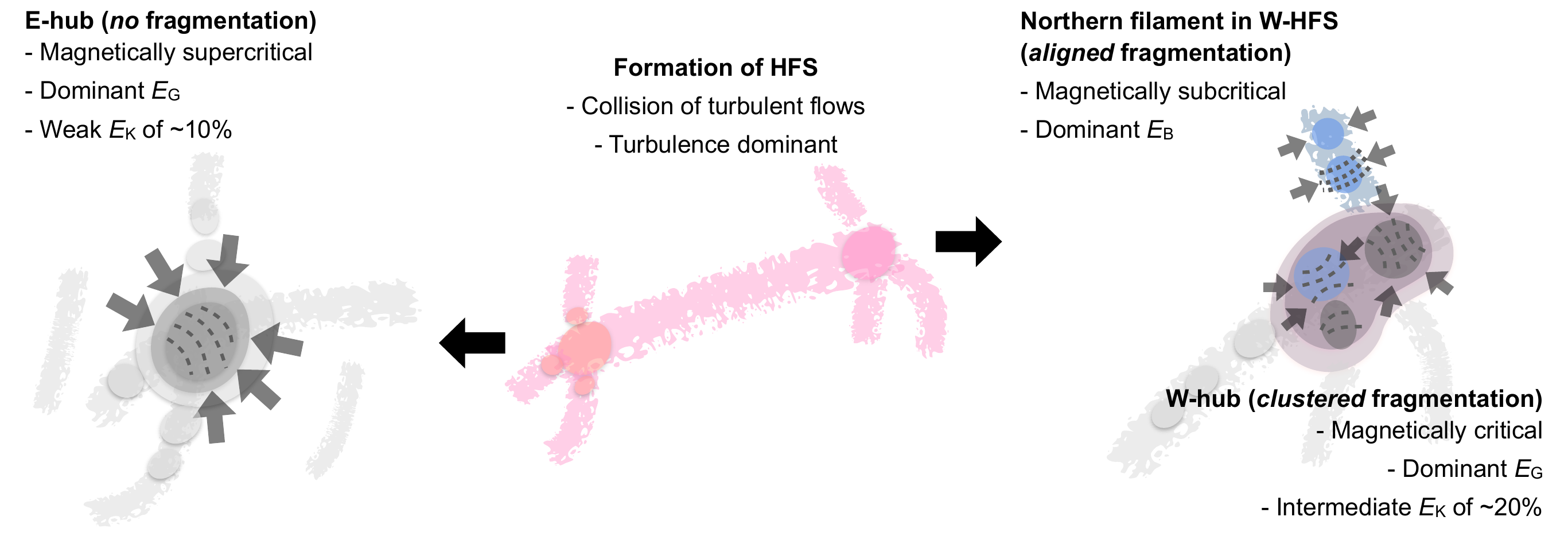}
\caption{Schematic for the roles of gravity, magnetic field, and turbulence in the formation and evolution of HFSs in IC~5146. {\it Center:} The initial formation of HFSs by collision of turbulent flows. {\it Left and Right:} The different evolution of hubs and filament into cores by fragmentation depending on different importance of $E_{\rm G}$, $E_{\rm K}$, and $E_{\rm B}$ in the E- and W-HFSs.  \label{fig:hubevol}}
\end{figure*}

In this section, we proposed a formation and evolution scenario of the E- and W-HFSs in the Streamer with what we found in our previous and present studies. 
As shown in Figure~\ref{fig:hubevol}, it is given as the turbulence driven stage at first and then fragmentation stage with its characteristic relative importance of gravity, magnetic field, and turbulence. 

At the turbulence driven stage, hubs are formed by the collision of turbulent flows. The colliding model is suggested in which filaments form by collision of turbulent flows and then the cores form in the turbulent dissipated stagnation point \citep{ballesteros1999,padoan2002}. \citet{chung2021} show that the E- and W-hubs in IC~5146 are supersonic while the cores forming in the two hubs are transonic which is consistent with the expectation of the colliding model. 

At the fragmentation stage, the different energy balances of gravity, magnetic field, and turbulence make the hubs to be differently fragmented. As discussed in the previous section, E-hub may have such dominant $E_{\rm G}$ but less $E_{\rm K}$ that it evolves into $no$ fragmentation. The small cores around E-47 may form at the very early stage of HFS when the turbulent flows collide, because the positions of small cores are matched with the overlapped regions of the E-hub and filaments (see the clumps and filaments in Figure~\ref{fig:ewhubs}). Hence, at the fragmentation stage, E-hub has not been fragmented any more but becomes a massive core, E-47. In the W-hub, turbulence is dissipated slowly and becomes important to disturb the regions in the hub and make a number of irregular cores. The turbulence of N-filament may have been quickly dissipated, making the filament to be possibly in a mode of the $aligned$ fragmentation type because of its dominant magnetic field energy. 

\begin{figure*} 
\includegraphics[width=1\textwidth,keepaspectratio]{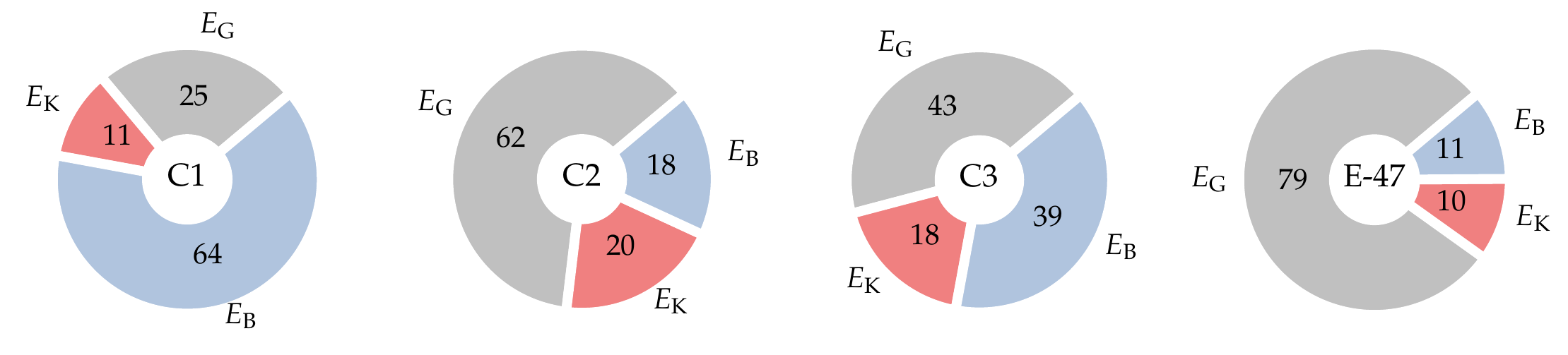} \caption{The relative importance of gravity, turbulence, and magnetic energies of C1, C2, C3, and E-47. The gravity energy ($E_{\rm G}$), kinematic energy ($E_{\rm K}$), and magnetic energy ($E_{\rm B}$) are presented with gray, red, and blue colors, respectively, and the relative portions are given in \%. \label{fig:energies2}}
\end{figure*}

We can expect how the cores evolve in the future with their energy balances and $|(2E_{\rm K}+E_{\rm B})/E_{\rm G}|$ values. The gravitational, kinematic, and magnetic field energies of cores are tabulated in Table~\ref{tab:bfield_hf}, and the relative portions of energies are presented in Figure~\ref{fig:energies2}. In all the cores, the dominant energy of each core is found to be consistent with that of their natal filament or hub, but the relative portions slightly differ from those of the filament or hub. In E-47, $E_{\rm G}$ has the largest portion like in E-hub. Moreover, E-47 has $|(2E_{\rm K}+E_{\rm B})/E_{\rm G}| = 0.4 \pm 0.2$, so it will continue to contract, maybe without further fragmentation. C2 has dominant $E_{\rm G}$ as W-hub has, and C2 and W-hub show almost the same relative portions of energies as well as $|(2E_{\rm K}+E_{\rm B})/E_{\rm G}|$ values (0.9$\pm$0.6 and 1.0$\pm$0.6, respectively), being close to the gravitational equilibrium. Hence, C2 is expected to fragment into smaller structures having $clustered$ distributions like its natal cloud, W-hub. C3 has slightly larger $|(2E_{\rm K}+E_{\rm B})/E_{\rm G}|$ than 1 (2$\pm$1). Though C3 has the largest portion of 43\% in $E_{\rm G}$, but it is hard to say $E_{\rm G}$ significantly dominates $E_{\rm B}$ whose portion is 39\%. C1 has dominant $E_{\rm B}$ (64\%) as N-filament has. In fact, the smallest $\delta \phi$ of C1 given in Table~\ref{tab:bfield_hf} already represented that $E_{\rm B}$ is dominant in C1. The $|(2E_{\rm K}+E_{\rm B})/E_{\rm G}|$ value of C1 is larger than 1 (3$\pm$2) and thus is believed to be in dispersion in future. According to the energy balance, C1 is expected to have smaller fragments lined up along the filament. \\ 

\section{Summary} \label{sec:summ}

To study the roles of the magnetic field, turbulence, and gravity in the evolution of HFS, we have carried out polarimetry and $\ceo~(3-2)$ observations with JCMT SCUBA-2/POL-2 and HARP toward the western HFS (W-HFS) of IC~5146. 
\begin{enumerate}
\item We identified a few tens of cores with 850~$\mu$m emission, and made analyses on the four 850~$\mu$m cores in the central $\sim 3^{\prime}$ area of observation with the best sensitivity. Among the four cores, one (C1) is on a northern filament (N-filament) of the W-HFS and three (C2, C3, and C4) are in the W-hub. The magnetic field geometry of C1 is perpendicular to the filament axis, while that in the W-hub is quite complex. The plane-of-the sky magnetic field strengths ($B_{\rm pos}$) are estimated for N-filament, W-hub, and cores. $B_{\rm pos}$ of C4 was not measured because the number of detected magnetic field vectors of C4 was not enough. The magnetic field strengths are in the range of $\sim$80 to 1200~$\mu$G. The mass-to-magnetic flux ratios are measured, finding that C1 is magnetically subcritical and C2 and C3 are marginally supercritical within the uncertainties. The Alfv\'enic Mach numbers estimated show that all the cores are in subsonic motions and have larger magnetic energy than the kinematic energy. 
\item We investigated the magnetic field morphologies of the W-hub and eastern hub (E-hub) to probe the edge-driven collapse and fragmentation scenario for the HFS formation in the Streamer. The magnetic field geometry of E-hub agrees to the expectation of the scenario, but that of W-hub does not. The curved B-fields of both hubs are supposed to be modified by the accretion flows and/or gravitational contraction in the hubs. 
The re-estimated magnetic field strength of E-47, the dominant central clump of E-hub, using the same method as the one for the cores in the W-HFS with the magnetic field angular dispersion and velocity dispersion given in W2019 is 500$\pm 100 ~\mu$G, being consistent to that of W2019. 
\item Referring to the suggestion for the fragmentation types of the HFS by \citet{tang2019}, we discussed the fragmentation scenario in E-HFS and W-HFS by using the relative importance among gravitational, kinematic, and magnetic energies ($E_{\rm G}$, $E_{\rm K}$, and $E_{\rm B}$) of the N-filament and W-hub as well as of E-hub. With the distribution of cores in them, we classified the N-filament of W-HFS into $aligned$, the W-hub into $clustered$, and E-hub into $no$ fragmentation, respectively. The relative portions of $E_{\rm G}$, $E_{\rm K}$, and $E_{\rm B}$ of filament and hubs are examined, and N-filament has dominant $E_{\rm B}$ (64\%) and E-hub has dominant $E_{\rm G}$ (76\%). This is well matched to the suggestion by \citet{tang2019}. W-hub is dominant in $E_{\rm G}$ (57\%), but it has relatively larger portion of $E_{\rm K}$ ($\sim$20\%) than E-hub has ($\sim$10\%). We argue that the slightly higher portion of kinematic energy might cause the $clustered$ fragmentation in the W-hub. 
\item We propose the evolutionary scenario of the E- and W-HFSs in the dark Streamer of IC~5146. From the turbulence properties of the HFSs and cores in them \citep{chung2021}, both of E- and W-HFSs have formed first by the collision of turbulent flows. In the E-hub, turbulence has been quickly dissipated and made $no$ fragmentation due to the dominant $E_{\rm G}$ and less $E_{\rm K}$. N-filament of W-HFS which has dominant $E_{\rm B}$ fragmented into $aligned$ cores. 
The current energy balance of W-hub indicates that the turbulence dissipation in the region has been slowly progressed, and turbulence has worked on disturbing the material, making W-hub to be in its $clustered$ fragmentation into C2, C3, and C4. From the current energy balances and the values of $|(2E_{\rm K}+E_{\rm B})/E_{\rm G}|$ of cores, it is expected that E-47 may continue contracting without any further fragmentation, and C1 and C2 may be fragmented into $aligned$ and $clustered$ substructures, respectively. \\
\end{enumerate}

The results support that the subtle different balance in gravitational, kinematic, and magnetic energy can cause the different type of fragmentation in a cloud proposed by \citet{tang2019}. We tentatively propose that, in the $E_{\rm G}$ dominant clouds, $\sim$20\% of $E_{\rm K}$ can make the material scattered and cause the $clustered$ fragmentation while $\sim$10\% $E_{\rm K}$ cannot disturb the material and hence make the cloud to be $no$ fragmentation. To improve the criteria of energy balance precisely, we will make more investigations toward various types of clouds and also with multi-scale observations. \\

\acknowledgments

The authors would like to acknowledge the anonymous referee for valuable comments that improved the quality of the paper. This work was supported by the National Research Foundation of Korea (NRF) grant funded by the Korea government (MSIT) (No. NRF-2019R1I1A1A01042480) and the National R \& D Program through the National Research Foundation of Korea Grants funded by the Korean Government (NRF-2016R1A5A1013277 and NRF-2016R1D1A1B02015014). E.J.C. is supported by Basic Science Research Program through the National Research Foundation of Korea(NRF) funded by the Ministry of Education(NRF-2022R1I1A1A01053862). C.W.L. is supported by the Basic Science Research Program through the National Research Foundation of Korea (NRF) funded by the Ministry of Education, Science and Technology (NRF-2019R1A2C1010851), and by the Korea Astronomy and Space Science Institute grant funded by the Korea government (MSIT) (Project No. 2022-1-840-05). W.K. was supported by the National Research Foundation of Korea (NRF) grant funded by the Korea government (MSIT) (NRF-2021R1F1A1061794). H.Y. was supported by Basic Science Research Program through the National Research Foundation of Korea(NRF) funded by the Ministry of Education (NRF-2021R1A6A3A01087238). 


\makeatletter
\renewcommand\@biblabel[1]{}
\makeatother

\end{document}